\documentclass{aastex}
\usepackage{emulateapj5}
\usepackage{apjfonts}

\makeatletter

\newenvironment{figurehere}
  {\def\@captype{figure}}
  {}
\makeatother

\newcommand{\csm}{C$_7^-$}

\newcommand{\cm}{cm$^{-1}$}

\slugcomment{Submitted to Astrophysical Journal Letters}

\shorttitle{\csm\ and the Diffuse Interstellar Bands}
\shortauthors{McCall et al.}

\begin{document}

\title{Rejection of the \csm\ Diffuse Interstellar Band Hypothesis}

\author{B.~J.~McCall, J.~Thorburn, L.~M.~Hobbs,
T.~Oka, and D.~G.~York}
\affil{Department of Astronomy \& Astrophysics, University of Chicago,
Chicago, IL 60637}
\email{bjmccall@fermi.uchicago.edu}

%% Notice that each of these authors has alternate affiliations, which
%% are identified by the \altaffilmark after each name.  Specify alternate
%% affiliation information with \altaffiltext, with one command per each
%% affiliation.

%\altaffiltext{1}{bjmccall@uchicago.edu}

\begin{abstract}

Using the new high resolution ($\sim$ 8 km/s)
echelle spectrograph on the 3.5-m telescope at the Apache Point Observatory,
we have begun a high sensitivity survey of the diffuse interstellar bands
in a large sample of
reddened stars.  Now that we are two years into this long-term survey,
our sample includes over 20 reddened stars which show at least one of the
DIBs that had been suggested to be caused by \csm, based on the gas phase
measurement of the \csm\ spectrum by J.~P.~Maier's group.

The high quality astronomical data from this larger sample of stars, along
with the spectroscopic constants from the new laboratory work recently
reported by Maier's group, have enabled us to examine more carefully the 
agreement between \csm\
and the DIBs.  We find that none of the \csm\ bands matches the DIBs in
wavelength or expected profile.  One of the DIBs ($\lambda$5748) attributed
to \csm\ is actually a stellar line.  The two strongest DIBs attributed to
\csm\ ($\lambda$6270 and $\lambda$4964) do not vary together in strength,
indicating that they do not share the same carrier.

On the whole, we find no evidence supporting the hypothesis that \csm\ is
a carrier of the diffuse interstellar bands.

\end{abstract}


\section{Introduction}

Perhaps the longest unsolved problem in astrophysical spectroscopy is that
of the Diffuse Interstellar Bands (DIBs), a series of hundreds of absorption
lines present in the visible spectra of nearly all reddened stars.  
It is now generally believed that the diffuse interstellar bands
are caused by free molecules in the gas phase \citep{herbig}, but 
despite many decades of effort by astronomers and molecular spectroscopists,
there has been no match between any subset of the diffuse bands and the
gas-phase laboratory spectrum of an individual molecule.

Many astronomers and molecular spectroscopists were hopeful that this impasse
had finally been broken when J.~P.~Maier's group reported 
\citep{tulej} a possible match between the gas-phase spectrum of \csm\ and five
DIBs in the catalog of \citet{jenniskens}.  The promising laboratory bands
are all vibronic bands of the lowest electronic transition 
($A^2\Pi_u \leftarrow X^2\Pi_g$) of \csm.  The strongest of the reported
bands, the origin ($0^0_0$) band at 6270.2 \AA, seemed to match the strong
$\lambda$6270 DIB.  The other four laboratory bands which seemed to match
the DIBs were the $1^1_0$ band at
5612.8 \AA\ (DIB at $\lambda$5610), $2^1_0$ at 5747.6 \AA\ ($\lambda$5748),
$3^1_0$ at 6063.8 \AA\ ($\lambda$6065), and the combination band $1^2_03^1_0$ 
at 4963 \AA\ ($\lambda$4964).

All five of these laboratory transitions seemed to agree with DIBs within
about 2 \AA, which is far closer agreement than had been achieved by any 
previously proposed DIB carrier.  Many of the astronomical observations of the
DIBs were at the limit of the sensitivity, as were the laboratory observations.
Because it was not possible to infer the rotational or spin-orbit constants
of \csm\ from the laboratory work, it was unclear how the bands might shift 
in wavelength or profile as a function of temperature.  For these reasons,
agreement within $\sim$2 \AA\ was sufficient to warrant further investigation.

Using initial data from our DIB survey \citep{myo}, we confirmed the
existence of four of the five DIBs, but had reservations about the
$\lambda$5748 band.  With data from four reddened stars, it appeared that
these four DIBs agreed reasonably well in both wavelength and 
relative intensities,
given the uncertainties in the laboratory data.  Additionally, in these
four sources (HD 46711, HD 50064, HD 183143, and Cygnus OB2 12) the four
bands seemed to vary together in intensity.

Recently, J.~P.~Maier's group has obtained laboratory data on the
$0^0_0$, $1^1_0$, $2^1_0$, and $3^1_0$ bands of \csm\ with
considerably higher resolution and sensitivity \citep{lakin}.  The authors
performed theoretical calculations to estimate the ground- and excited-state 
rotational and spin-orbit constants, and then varied the spin-orbit
constants to best fit their experimental spectrum.  Since the overall profile
of the spectrum is very different as the spin-orbit constants are varied,
this approach results in a fairly unambiguous determination of the molecular
constants (though not as unambiguous as would be possible from a fully
rotationally-resolved spectrum).  With the constants determined from the
experiment,
it is now possible to predict how the 
\csm\ spectrum will change with temperature.  Such predictions are essential
in performing a detailed comparison with the DIBs.

At the same time, our DIB survey has progressed to the point where
we now observe at least some of the bands attributed to \csm\ in the
spectra of over twenty reddened stars.  Additionally, our data reduction
pipeline has improved substantially, such that the aliasing which limited
the signal-to-noise in our earlier work has been completely eliminated.  These
advances in both the laboratory and astronomical spectroscopy have prompted
us to re-examine the case for \csm\ as a diffuse band carrier.

\section{Observations and Data Reduction}

The observations reported here are part of our long-term survey of the
DIBs in a large sample of stars.  High resolution ($R \sim$ 37,500) 
visible (4000--10000 \AA) spectra have been obtained with the Astrophysical
Research Consortium Echelle Spectrograph (ARCES) on the 3.5-m telescope 
at the Apache Point Observatory.  Data reduction is performed using
standard IRAF routines, as described in detail by \citet{thorburn}.
A more complete description of our DIB survey will be given in a future paper.

\section{Results and Discussion}

\subsection{Simulation of \csm\ spectra}

Given the constants from \citet{lakin}
($B''$=897 MHz, $B'$=887 MHz, $A_{SO}''$=27.4 \cm, and $A_{SO}'$=0.6 \cm),
we used the method of \citet{vanvleck} to calculate the energy levels of
\csm\ and the intensity factors for the individual rotational lines
within a given vibronic band.  [We assumed the same constants for each
vibronic band, as the vibrational dependence of the constants is expected
to be smaller than the uncertainty in the determined constants.]
The populations of the
rotational states of \csm\ were then calculated using a Boltzmann expression
at a given temperature, from which we were able to simulate the absorption
spectrum.  Temperatures between 10 K and 90 K were considered, as diffuse
clouds can be expected to have temperatures within this range.
For the linewidth of each transition, we assumed a
Gaussian profile with full-width at half-maximum 10 km/s, which is the 
FWHM of the observed K \textsc{i} lines in HD 185418 and HD 229059 (two
stars we have chosen for the comparsion due to their narrow K \textsc{i}
lines).

\begin{figurehere}
\begin{center}
\scalebox{0.8}{\includegraphics{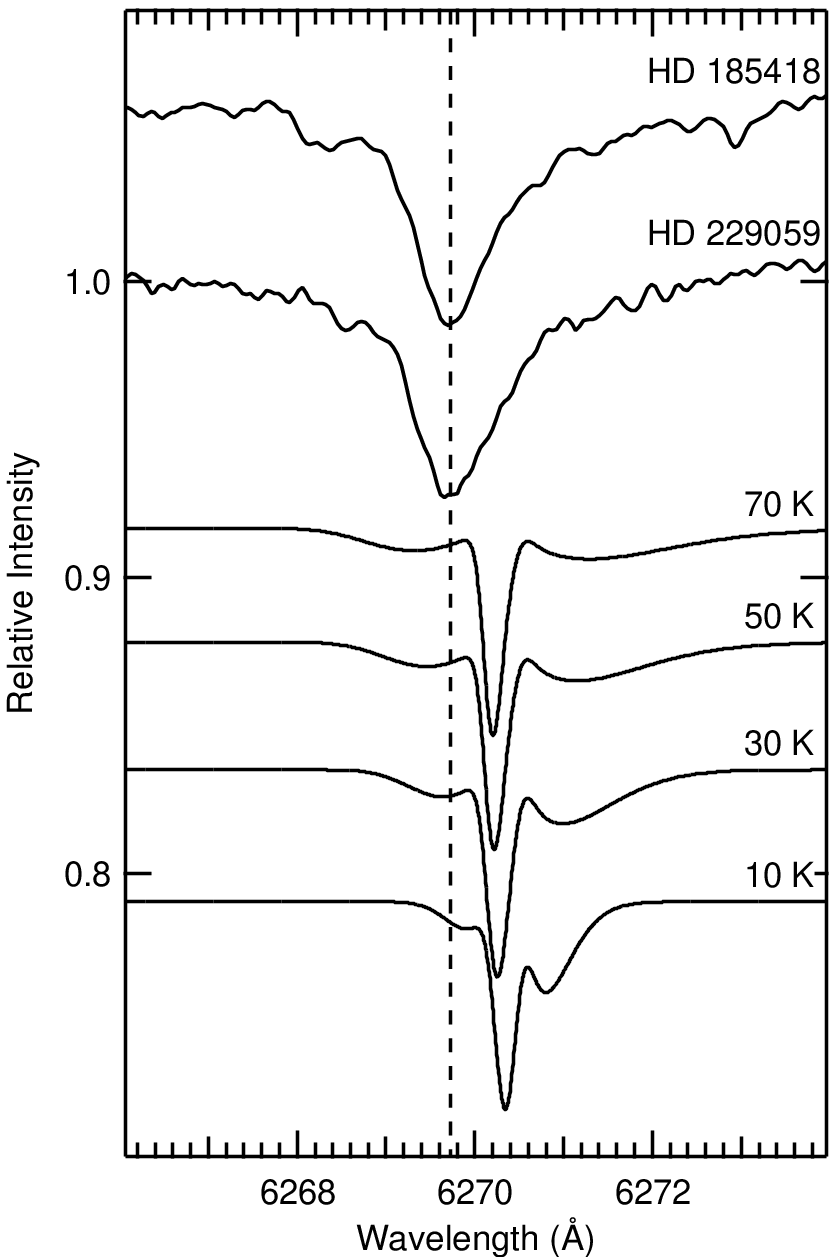}}
\caption{Spectra of the $\lambda$6270 DIB in two reddened stars (upper
traces), compared with simulations of the $\Omega''$=1/2 component of
the $A \leftarrow X$ $0^0_0$ origin band of \csm\ at
various temperatures.  The simulations assume a Gaussian linewidth of
10 km/s, derived from the K I $\lambda$7699 line (not pictured).
Note the lack of agreement between \csm\ and the DIB, both in wavelength
and in profile.
\label{origin}}
\end{center}
\end{figurehere}

\subsection{Comparsion between DIBs and simulated \csm\ spectra}

We begin by considering the $\Omega''$=1/2 spin-orbit component of the 
origin ($0^0_0$) band of \csm, in comparison with the $\lambda$6270 DIB.
The origin band is naturally the strongest of the laboratory features, and
$\lambda$6270 is also by far the strongest of the DIBs suggested to
correspond to \csm.  Figure \ref{origin} shows the spectra of $\lambda$6270
toward HD 185418 and HD 229059, along with the simulations of the \csm\
origin band at temperatures of 10, 30, 50, and 70 K.  As can be seen from
the figure, neither the central wavelengths nor the profiles of the \csm\ 
spectra agree with the $\lambda$6270 diffuse band.  This disagreement
argues strongly against the assignment of $\lambda$6270 to \csm.

In Figure \ref{originsat} we consider both the $\Omega''$=1/2 (left) and
$\Omega''$=3/2 (right) components of the \csm\ origin band.  Because
$\Omega''$=3/2 is higher in energy, the intensity of the right-hand component
increases with temperature, as evident in the simulations at 30, 60, and 90 K.
In Figure \ref{originsat}, an (unreasonably large) {\em ad hoc} Gaussian
linewidth of 30 km/s has been assumed in order to improve the agreement
with $\lambda$6270.  It is difficult to state with certainty because of
the presence of the strong $\lambda$6284 DIB, but it appears
that there is little evidence for the $\Omega''$=3/2 component in the
astronomical spectra.
However, because the intrinsic profile of the strong $\lambda$6284 DIB is
not known, the presence of the $\Omega''$=3/2 component cannot be 
definitively ruled out.

\begin{figurehere}
\begin{center}
\scalebox{0.8}{\includegraphics{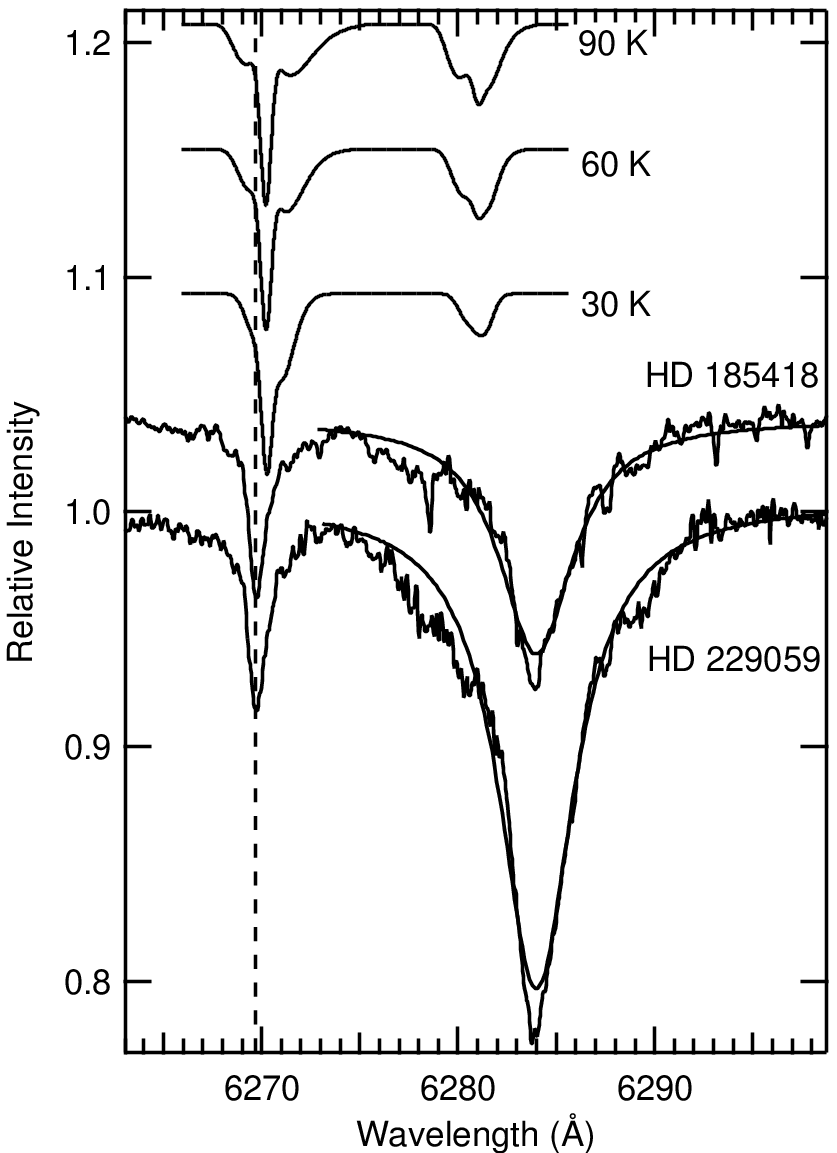}}
\caption{Simulations of both components ($\Omega''$=1/2 on the left
and $\Omega''$=3/2 on the right) of
the \csm\ origin band (upper traces).  In this figure, the simulations were
performed assuming an (unreasonably large)
{\em ad hoc} linewidth of 30 km/s in order to better
match the width of $\lambda$6270 for comparison.  The lower traces show the 
spectra of HD 185418 and HD 229059.  These spectra have been divided by
standard stars (HD 149757 and HD 229059, respectively) in order to remove
atmospheric absorption lines of O$_2$.  The smooth 
curves are Lorentzian fits to the $\lambda$6284 DIB.  See the text for
a discussion of the $\Omega''$=3/2 component.
\label{originsat}}
\end{center}
\end{figurehere}

Figure \ref{band110} compares the simulated spectrum of the $1^1_0$
vibronic band of \csm\ to the $\lambda$5610 DIB.  In this case, the
wavelength discrepancy between the \csm\ band and the DIB is particularly
egregious, over 2 \AA.  In addition, the profile is considerably different ---
the simulated spectrum shows a sharp band-head, while the DIB has a fairly
Gaussian profile.  There is no reason to attribute the $\lambda$5610 DIB
to \csm, and no evidence for any astronomical feature resembling the
$1^1_0$ band of \csm.

\begin{figurehere}
\begin{center}
\scalebox{0.75}{\includegraphics{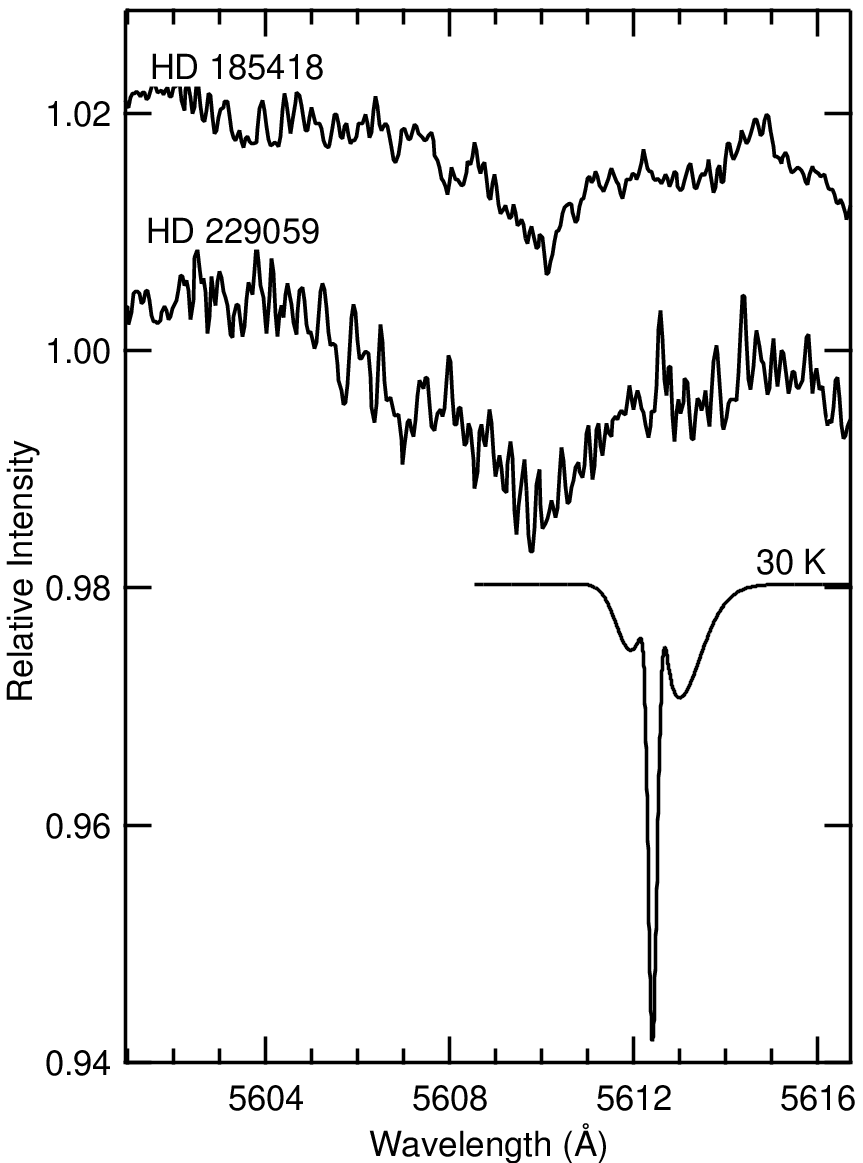}}
\caption{Spectra of the $\lambda$5610 DIB in HD 185418 and HD 229059,
compared with a simulation (10 km/s linewidth) of the $\Omega''$=1/2 component 
of the $1^1_0$ band of \csm\ at 30 K.  Note the disagreement in wavelength
and profile between \csm\ and the DIB.
\label{band110}}
\end{center}
\end{figurehere}

\begin{figurehere}
\begin{center}
\scalebox{0.75}{\includegraphics{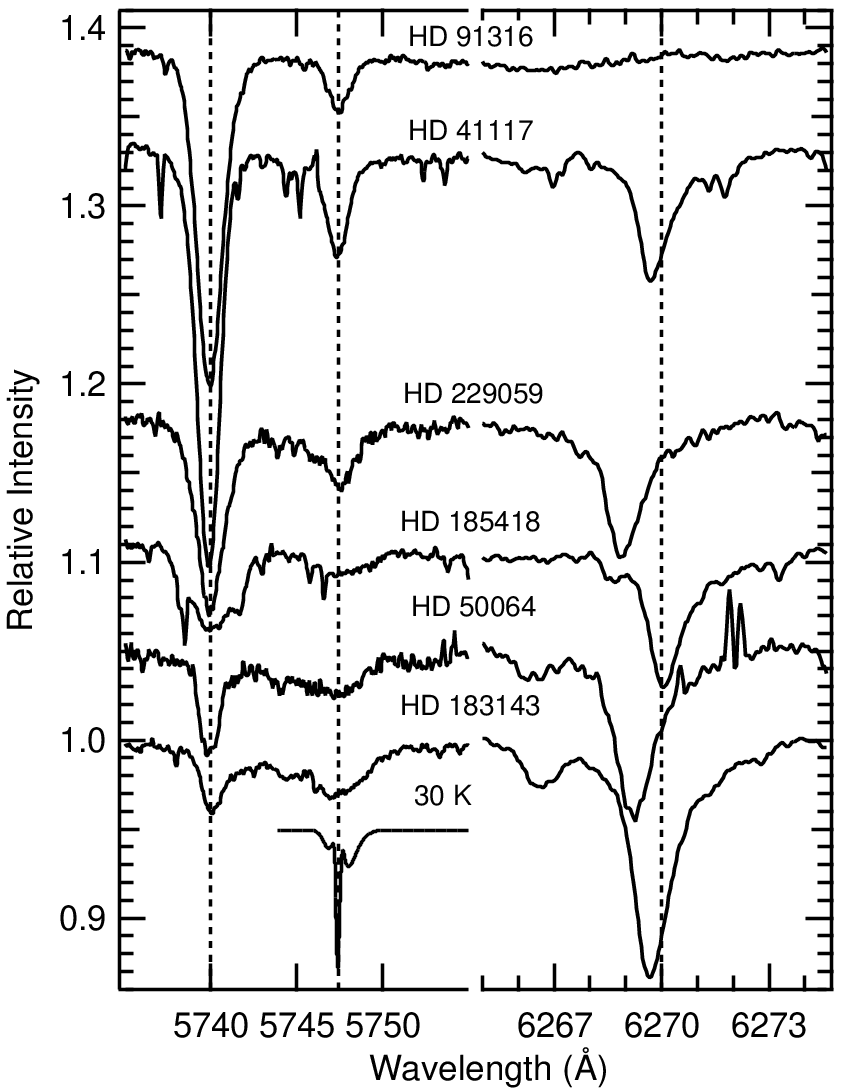}}
\caption{Spectra of the region near 5747 \AA\ (left) and 6270 \AA\ (right)
in one unreddened star (HD 91316) and several reddened stars.  The spectra
have been shifted in wavelength to align the Si \textsc{iii} stellar line 
at 5740 \AA.  Note that the feature at 5747 \AA\ now has the same wavelength
from star to star, in contrast to $\lambda$6270.  This, along with the fact
that the 5747 \AA\ feature is seen in the unreddened star HD 91316 where
the diffuse bands are absent, shows that the 5747 \AA\ line is a stellar
feature rather than a DIB, and only $\lambda$6270 is of interstellar origin.
For reference, a simulation of the \csm\
$2^1_0$ band (10 km/s linewidth) is also displayed.
\label{band210}}
\end{center}
\end{figurehere}

Figure \ref{band210} shows the region where the $2^1_0$ band of \csm\ is
expected, as well as the $\lambda$6270 DIB (which has been
suggested to correspond to the origin band).  In this
figure, the spectra have been shifted in wavelength in order to co-align
the Si \textsc{iii} stellar line at 5740 \AA.  It is easily seen from the
figure that with this wavelength shift, the feature at 5747 \AA\ is also
aligned, whereas the diffuse interstellar band $\lambda$6270 is no longer
aligned.  This implies that the feature which \citet{jenniskens} claim
as a ``certain'' DIB at 5748 \AA\ is, in fact, a stellar line.  This is
particularly clear from the strength of the feature in the unreddened star
HD 91316 ($\rho$ Leo) which shows no trace of the $\lambda$6270 DIB.
Since ``$\lambda$5748'' is not of interstellar origin, it cannot be
assigned to \csm.

Figure \ref{band310} examines the case of the $3^1_0$ band of \csm,
compared with the $\lambda$6065 DIB.  Here we see that there is again
a pronounced wavelength discrepancy $\stackrel{>}{_\sim}$ 1 \AA\ between
\csm\ and the DIB.
Once again, there is no evidence to support assigning $\lambda$6065
to \csm.  [It is interesting to note that in our present sample of stars,
$\lambda$6065 and $\lambda$6270 appear to be correlated in intensity.  Thus,
while these bands are probably not due to \csm, they may share a common
or closely (chemically) related carrier.]

\begin{figurehere}
\begin{center}
\scalebox{0.8}{\includegraphics{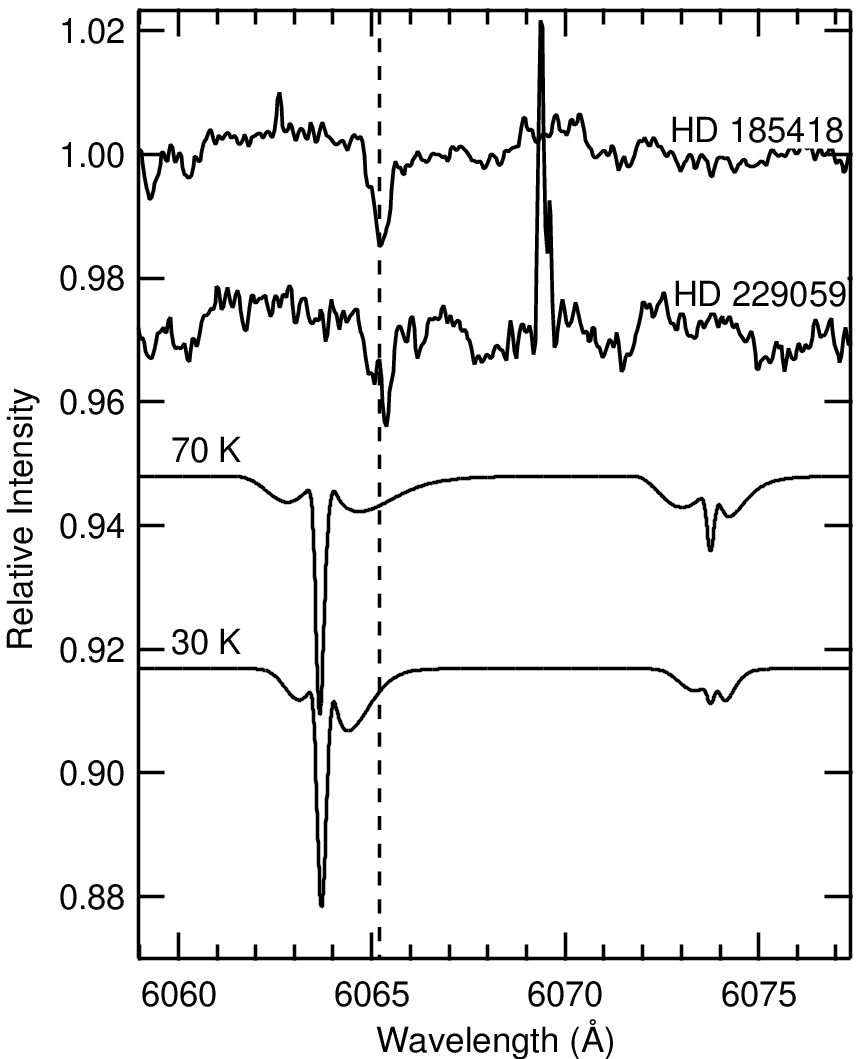}}
\caption{Spectra of the region near $\lambda$6065 in HD 185418 and HD 229059,
along with simulations (10 km/s linewidth) of the \csm\ $3^1_0$ band at 30 K 
and 70 K.  Note the poor wavelength agreement between \csm\ and the DIB.
\label{band310}}
\end{center}
\end{figurehere}

\subsection{Other bands of \csm}

The combination band $1^2_03^1_0$ is surprisingly strong in the laboratory
spectrum of \citet{tulej}, and it was suggested that this band may
correspond to the $\lambda$4964 DIB.  Since the $1^2_03^1_0$ band was not 
revisited in the experiment of \citet{lakin}, we cannot examine in detail 
its agreement with the $\lambda$4964 DIB.  However, with our substantially
larger sample of stars, we are in a position to re-examine the correlation
between the intensities of $\lambda$4964 and $\lambda$6270 (supposedly the
origin band of \csm).  If these two bands are due to the same species, they
must have the same intensity ratio from star to star, as this ratio is
determined solely by the Franck-Condon factors.

\begin{figurehere}
\begin{center}
\scalebox{0.8}{\includegraphics{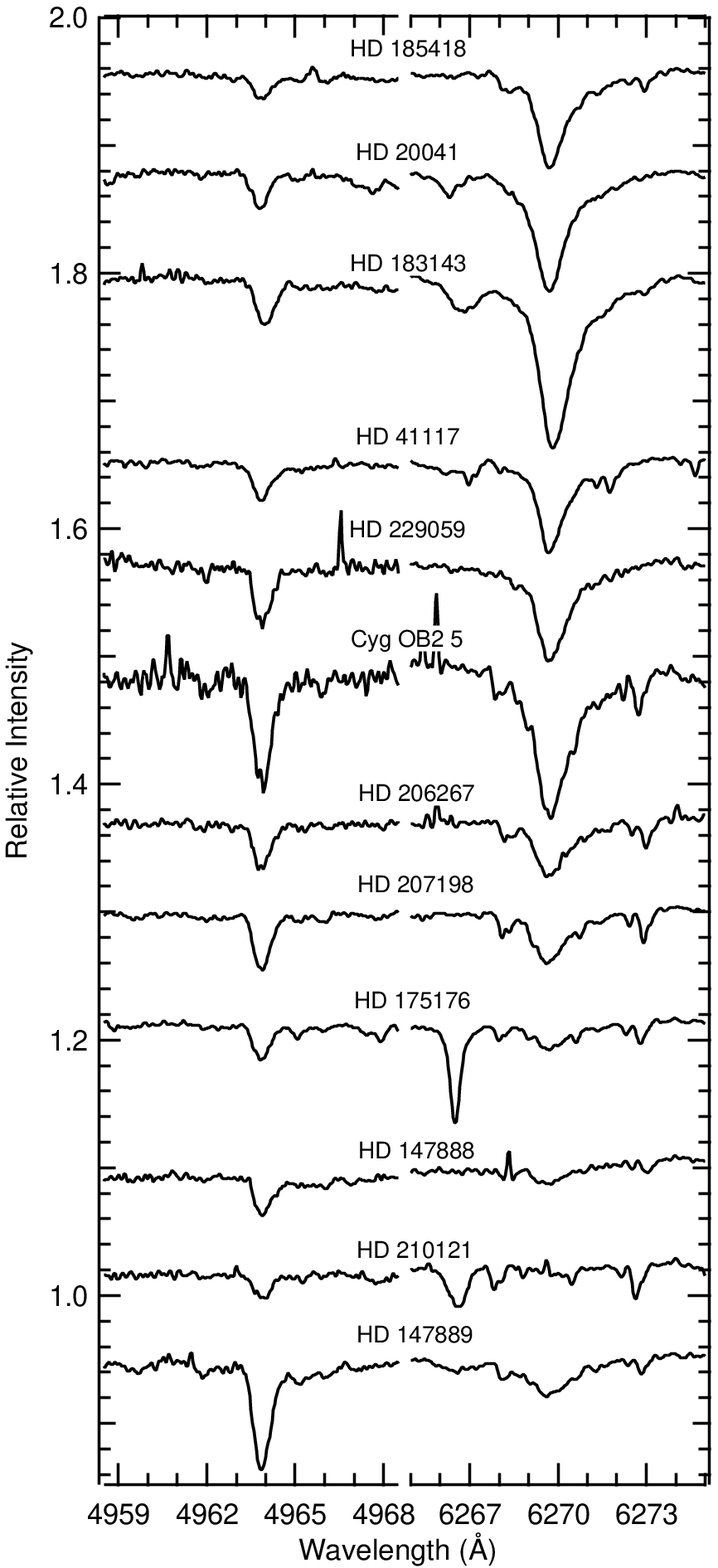}}
\caption{Spectra of the $\lambda$4964 (previously attributed to \csm\
$1^2_03^1_0$) and $\lambda$6270 (\csm\ $0^0_0$) DIBs in several
reddened stars.  Note the lack of correlation between the intensities of
the two bands, indicating that they do not have a common carrier.
\label{band120310}}
\end{center}
\end{figurehere}

Figure \ref{band120310} displays the spectra of $\lambda$4964 and
$\lambda$6270 in a sample of twelve reddened stars.  While it appeared
in our original work \citep{myo} that these bands were correlated, this
was apparently due to the small sample (4) of stars considered in that work.
From this figure it is evident that in some stars (e.g.\ HD 183143 and
HD 20041) $\lambda$6270 is much stronger than $\lambda$4964, while in 
other stars (e.g.\ HD 147888 and HD 147889) the situation is reversed.
This clearly rules out the possibility that both bands can be due to the
same carrier, and therefore they cannot both be due to \csm.

There are two other weak vibronic bands of the $A \leftarrow X$ transition
of \csm\ that were reported by \citet{tulej}.  These both happen to be
doublets: $1^2_0$ at 5089.5 and 5095.7 \AA, and $1^1_03^1_0$ at
5449.6 and 5456.7 \AA.  We were not able to detect these bands in our
astronomical spectra, but because of the intrinsic weakness of these bands
(compared with the origin band) we were not able to set useful upper limits
on them either.  Similarly, we were not able to obtain a useful limit for
the origin band of the $B \leftarrow X$ band, which has a very small central
depth due to its intrinsic broadness.

\section{Conclusions}

The hypothesis that \csm\ is a diffuse interstellar band carrier has
been very attractive on spectroscopic grounds alone --- no previously
proposed carrier has come so close to providing a wavelength match to any
set of the diffuse bands.  There are strong chemical arguments against this
hypothesis: chemical models \citep{ruffle} are unable to reproduce the
necessary abundance of \csm, even with the most favorable 
assumptions.  This is due in large part to the destruction of \csm\
by hydrogen atoms, which has been recently confirmed to proceed with a
fast rate coefficient \citep{bierbaum}.
In spite of these chemical arguments, the approximate coincidence between
the \csm\ and DIB wavelengths has been too close to ignore, given the 
uncertainties inherent in the previously available laboratory and
astronomical work.  

Armed with the spectroscopic constants of \csm\ from \citet{lakin} and
our improved sample of DIB observations, however, it is now clear that \csm\
fails the stringent tests enabled by high resolution spectroscopy.  The
origin band does not match $\lambda$6270 in wavelength or profile, and there
is no sign of the higher-lying $\Omega''$=3/2 component.  The $1^1_0$ band
is way off in wavelength from $\lambda$5610 ($\sim$ 2 \AA) and also does not 
agree with the profile of the DIB.  The DIB attributed to the $2^1_0$ band
turns out to be a stellar line.  The $3^1_0$ band does not match 
$\lambda$6065 in wavelength or profile.  Finally, the DIBs attributed to
the $1^2_03^1_0$ band ($\lambda$4964) and the origin band ($\lambda$6270)
do not vary together in intensity, and therefore do not share a common
carrier.

Close as the wavelength match appeared to be at first sight, there now seems
to be no evidence to support the hypothesis that \csm\ is a carrier of the
diffuse interstellar bands.

\acknowledgements

The authors thank J.~P.~Maier and his research group for providing their
laboratory data in advance of publication.
This work has made use of the NASA Astrophysics Data Service, as well as
the SIMBAD database at the Centre de Donn\'{e}es astronomiques de Strasbourg.
This work has been supported by NSF grant PHYS-9722691 and 
NASA grant NAG5-4070.  B.J.M.\ has been supported by
the Fannie and John Hertz Foundation.

%clearpage

\clearpage

%%%%%%%%%% FIGURES %%%%%%%%%%%%
%
%\begin{multicols}{2}
%
%
%\end{multicols}
%
%
%%%%%%%%%% TABLES %%%%%%%%%%%%
%
%

\end{document}